\documentclass[aps,prd,twocolumn,nofootinbib]{revtex4-2}

\usepackage{amsmath,amssymb,graphicx}
\usepackage{hyperref}

\begin{document}

\title{Model-Independent Bound on Neutrino Energy Reconstruction from Nuclear Targets}

\author{Sanjeev Kumar Verma}
\email{skverma@physics.du.ac.in}
\affiliation{Department of Physics and Astrophysics, University of Delhi, Delhi, INDIA - 110007}

\date{\today}

\begin{abstract}
Neutrino energy reconstruction on nuclear targets underlies oscillation measurements and precision tests of weak interactions. Inclusive charged–current data have long exhibited degeneracies commonly attributed to axial-mass tuning, multinucleon dynamics, and final-state interactions. This work shows that, even in the idealized limit of perfect detectors and exact nuclear dynamics, inclusive lepton-only reconstruction admits no unique inverse. A strictly positive lower bound on neutrino energy resolution follows from the finite energy-transfer support of the inclusive nuclear axial response. The result identifies an irreducible contribution to reconstruction uncertainty that is independent of modeling assumptions and experimental resolution.
\end{abstract}

\maketitle

\section{Motivation and Scope}

Long-baseline neutrino oscillation experiments infer the incident
neutrino energy from interactions on nuclear targets. In charged--current
measurements, this inference is typically based on the kinematics of the
outgoing charged lepton.

It is well established that distinct nuclear descriptions---including
variations of the axial mass, the treatment of multinucleon excitations,
and alternative prescriptions for final-state interactions---can
reproduce the same inclusive lepton spectra while implying different
neutrino energy reconstructions. Such degeneracies have been documented
across a wide range of theoretical and experimental studies and are
commonly regarded as a limitation of nuclear modeling
\cite{Formaggio,Martini}.

A complementary line of work has examined the resulting reconstruction
biases within specific nuclear models and reconstruction prescriptions,
demonstrating that identical lepton kinematics may correspond to
different true neutrino energies \cite{Ankowski,Martini,Formaggio}. These studies quantify
the phenomenological impact of the ambiguity.

The analysis below addresses a logically prior question. Even if nuclear
dynamics were known exactly and experimental resolution were ideal,
would inclusive charged--current data permit unique neutrino energy
reconstruction from lepton-only observables? The answer is negative.
Inclusive measurements retain a nonzero ambiguity arising from
structural non-identifiability, independent of modeling assumptions and
detector effects. This ambiguity admits a model-independent lower bound
set by the inclusive nuclear axial response.

\section{Inclusive Formalism and Axial Response}

Consider the inclusive charged--current process
\begin{equation}
\nu_\ell(k) + A \rightarrow \ell^-(k') + X ,
\end{equation}
where $\nu_\ell$ is an incident neutrino of flavor $\ell$ with
four-momentum $k^\mu=(E_\nu,\mathbf{k})$, $\ell^-$ is the outgoing charged
lepton with four-momentum $k'^\mu=(E_\ell,\mathbf{k}')$, and $X$ denotes
an arbitrary hadronic final state.

The inclusive differential cross section is written as
\begin{equation}
\frac{d\sigma}{dE_\ell d\Omega_\ell}
= \frac{G_F^2 \cos^2\theta_C}{4\pi^2}
\, L_{\mu\nu}(k,k') \, W^{\mu\nu}(q),
\end{equation}
where $G_F$ is the Fermi constant, $\theta_C$ is the Cabibbo angle,
$L_{\mu\nu}$ is the leptonic tensor, and $W^{\mu\nu}$ is the inclusive
hadronic tensor. The four-momentum transfer is
$q^\mu = k^\mu - k'^\mu = (\omega,\mathbf{q})$, with invariant
$Q^2 \equiv -q^2$. This response-function formulation follows the
standard inclusive treatment developed in electron--nucleus scattering
\cite{Donnelly,Benhar}.

The axial contribution may be isolated by introducing a single inclusive
axial response functional
\begin{equation}
\mathcal R_A(Q^2,\omega),
\end{equation}
which serves as an operational representative of the nuclear axial
response after summing over all hadronic final states. Response
functions of this type are standard in inclusive lepton--nucleus
scattering, where axial strength is distributed over energy transfer
and momentum \cite{Donnelly,Benhar}. Here, these contributions are
packaged into a single scalar functional, defined implicitly by its
role in the inclusive cross section once all unobserved hadronic
degrees of freedom have been integrated out. In particular, only the
outgoing lepton is observed, while the energy transfer $\omega$ is not
measured and is integrated over in the cross section.

Because the energy transfer $\omega$ is unobserved, inclusive
charged--current measurements retain no event-by-event information
beyond that encoded in the functional dependence of
$\mathcal R_A(Q^2,\omega)$. All microscopic details of the nuclear
dynamics enter the observable cross section only through this
inclusive response. As a result, any statement about the possibility
of neutrino energy reconstruction from lepton-only observables must be
formulated directly at the level of $\mathcal R_A$. The following
section uses this observation to establish a model-independent bound
on neutrino energy resolution.

\section{Inclusive Axial Response Bound}

\textbf{Theorem.}—
For an inclusive charged--current measurement on a nuclear target, the
neutrino energy reconstructed from lepton-only kinematics satisfies
\begin{equation}
\Delta E_\nu \ge \frac{1}{2}
(\omega_{\max}-\omega_{\min}),
\label{eq:bound}
\end{equation}
where $[\omega_{\min},\omega_{\max}]$ denotes the support of
$\mathcal R_A(Q^2,\omega)$ at fixed measured lepton kinematics.

\emph{Derivation.}—
For fixed outgoing lepton energy $E_\ell$ and solid angle $\Omega_\ell$,
the inclusive differential cross section may be written in the form
\begin{equation}
\frac{d\sigma}{dE_\ell d\Omega_\ell}
= \int_{\omega_{\min}}^{\omega_{\max}} d\omega \;
K(E_\ell,\Omega_\ell;\omega)\,
\mathcal R_A(Q^2,\omega),
\end{equation}
where $K$ is a leptonic kernel determined entirely by electroweak
kinematics and fixed by the measured lepton variables. Its explicit
form is immaterial for the present argument. Equation~(5) is to be read
as the formal definition of $\mathcal R_A(Q^2,\omega)$: it is the unique
scalar function of the energy transfer whose convolution with the
known kernel reproduces the inclusive cross section once all hadronic
degrees of freedom are summed over. All dependence on unobserved
nuclear dynamics enters solely through $\mathcal R_A$.

Energy conservation relates the incident neutrino energy to the
(unobserved) energy transfer,
\begin{equation}
E_\nu = E_\ell + \omega .
\end{equation}
Because $\omega$ is not measured, a fixed lepton configuration
$(E_\ell,\Omega_\ell)$ is compatible with a set of neutrino energies
\begin{equation}
\mathcal E_\nu
= \{\, E_\ell + \omega \;|\; \omega \in
[\omega_{\min},\omega_{\max}] \,\}.
\end{equation}
Because $\omega$ is unobserved, Eq.~(6) does not determine a unique
value of $E_\nu$. Instead, the measurement is compatible with a set of
neutrino energies, denoted $\mathcal E_\nu$, whose elements satisfy
Eq.~(6).

The inverse problem of neutrino energy reconstruction is therefore
\emph{set-valued}, in the sense that a single measured lepton
configuration is compatible with a range of incident neutrino energies
rather than a unique value. Any estimator based solely on inclusive
lepton kinematics must then assign a reconstructed energy lying within
the \emph{convex hull} of $\mathcal E_\nu$, i.e., within the smallest
interval containing all neutrino energies consistent with the
measurement. The diameter of this hull is
\begin{equation}
\mathrm{diam}(\mathcal E_\nu)
= \omega_{\max}-\omega_{\min}.
\end{equation}
The minimal achievable resolution is bounded from below by half this
width, yielding Eq.~(\ref{eq:bound}). \hfill$\square$

\section{Illustration with Physical Scales}

Inclusive electron--nucleus scattering data show that, at fixed momentum
transfer, nuclear response strength is distributed over a broad range of
energy transfer. For light and medium nuclei such as C, O, and Ar, the
support of the inclusive response typically spans
\begin{equation}
\omega_{\max}-\omega_{\min}
\sim \mathcal O(100~\mathrm{MeV}),
\end{equation}
as established in inclusive $(e,e')$ measurements and response-function
analyses \cite{Donnelly,Benhar}. Superscaling studies provide a compact
representation of this empirical width without introducing additional
model assumptions \cite{Amaro}. When inserted into
Eq.~(\ref{eq:bound}), this scale implies an irreducible neutrino-energy
uncertainty at the level of several tens of MeV.

This scale is smaller than the effective energy resolutions reported in
realistic oscillation analyses, which include additional smearing from
detector effects, event selection, and flux convolution. The bound
therefore represents a minimal contribution that persists even in the
idealized limit of perfect measurements.

\section{Implications}

Distinct microscopic mechanisms---including axial-mass variation,
multinucleon strength, and final-state interactions---correspond to
different parameterizations of the same inclusive functional
$\mathcal R_A$. Comparative model studies have shown that such
mechanisms can yield nearly identical inclusive lepton spectra while
differing in the inferred neutrino energy \cite{Ankowski,Martini}. From
this perspective, model-dependent analyses quantify the magnitude of an
ambiguity whose existence follows more generally from the
non-invertibility of inclusive lepton-only kinematics.

The result does not constrain measurements incorporating exclusive
hadronic information or additional kinematic tagging. Within an
inclusive framework, however, no refinement of nuclear modeling can
remove the structural ambiguity implied by the finite support of the
axial response.

\section{Conclusion}

Inclusive charged--current measurements on nuclear targets admit a
model-independent lower bound on neutrino energy reconstruction from
lepton-only observables. The bound reflects structural
non-identifiability rather than deficiencies of nuclear modeling or
experimental resolution. Restoration of sufficient exclusive hadronic
information is required to evade this limitation.


\end{document}